# The Search for the Cosmological Axion - A New Refined Narrow Mass Window and Detection Scheme

Masroor H. S. Bukhari*,
*Department of Physics, Faculty of Science,
Jazan University, Jazan 45142,
Saudi Arabia.*

## Abstract

Based upon a previous axion mass proposal and detection scheme, as well as considering the axion mass ranges suggested by persuasive simulations in recent years, we present a revised axion/ALP search strategy and our calculations, concentrating on a slightly narrower axion mass (and corresponding Compton frequency) window in this report. The sole aim of this study was to provide a pin-point mass value for a resonant cavity detection scheme, and the outcome of calculations presented here reinforces the earlier mass window and the results (slightly different values from our calculations but within the same window) obtained by the calculations and simulations by Kawasaki *et al.* (2015) and Buschmann *et al.* (2020). The window comprises the spectral region of 18.99 to 19.01GHz (that falls within the Ku microwave band), with a center frequency of 19.00GHz ($\pm$0.1GHz), with equivalence to an axion mass range of 78.6 to 79.6 μeV, with the center mass at the value of 78.582 ($\pm$5.0) μeV, our suggested most likely value for an axionic/ALP field mass, if these fields exist. Our search strategy, as summarized herewith, is based upon the assumption that the dark matter that exists in the current epoch of our physical universe is dominated by axions and thus the local observable axion density is the density of the light cold dark matter, permeating our local neighborhood (mainly in the Milky Way galactic halo). Some ideas and the design of an experiment, based upon the *inverse Primakoff effect*, and built around a Josephson Parametric Amplifier and Resonant Tunneling Diode combination installed in a resonant RF cavity, are also some possibly novel ideas, as introduced in this report.

* Email: mbukhari@jazanu.edu.sa
ORCID ID: https://orcid.org/0000-0003-3604-3152

## 1. Introduction

In order to understand the axion mass problem it is necessary to get acquainted with the phenomenology of the elusive particles known as *"Axions"* or *"Axion-like Particles (ALPs)"*, which seem to lie at the boundary of or beyond the current Standard Model (SM) of physics. These particles with interesting properties and possible liaisons with the current model of cosmology (the "Concordance Model" or "Λ Cold Dark Matter" model) have great potential in understanding the nature of matter (and possibly the nature of *"dark matter"* as well.)

A deep analysis of the theory of Quantum Chromodynamics (QCD), one of the cornerstone constituents of the SM Lagrangian, revealed that a problem existed with the vacuum structure in the theory (specifically by the virtue of the vacuum angle's or ($\overline{\Theta}$) parameter's extremely small value, also the minimum of the QCD Instanton potential, which is estimated at around less than $10^{-10}$ from the absence of neutron electric dipole moment) which allowed for the violation of Charge Conjugation and Parity (CP) symmetries in QCD interactions, a dilemma which was later termed as the *"Strong CP"* problem [1]. Physicists R. Peccei and H. Quinn suggested an elegant mechanism to fix this problem with the suggestion of a new chiral symmetry ($U(1)_{PQ}$ [2], which spontaneously broke to result into the emergence of an extremely light, virtually invisible, pseudo-Nambu-goldstone boson, named as the *Axion* (although lately its variants have also been suggested, known as *"Axion-Like Particles"* or *ALP's*). These were also named and classified as the *"Invisible Dark Matter"* particles by investigators during the early stages of the theory [3].

Axionic field replaced the small vacuum angle and emerged as a new quantum field among all the possible fields existing within the quantum vacuum, however with extremely weak coupling to ordinary matter. It did not take long before axions were identified as plausible new candidates for cold dark matter dominating the matter density of our universe [4, 5]. There has been a lot of theoretical investigation into the physics of axions and countless models for their existence and properties have been suggested, including the recent suggestions of coherent zero-mode axionic oscillations around their minimum, possibly forming a Bose-Einstein Condensate (BEC) of axions [6, 7].

Within the framework of QCD and within the string theoretical framework, some ideas have been developed during the past three decades in an attempt to understand the formation and evolution of axions soon after the onset of inflation during the epoch. The detailed treatment of the primordial axion formation from the string networks is beyond the scope of this report and can be found in any detailed reviews, such as in [8] and [9]. In short, during the Planck epoch, soon after the PQ symmetry breaking, the fundamental axionic string networks form and give way to the formation of string-based domain walls which interact and annihilate around the time of QCD phase transition [8, 9, 10, 11, 12], in turn producing axions which permeate the universe and reach the present epoch as *non-thermal Relic Axions* [13]. In the beginning these axions are hot and have a much higher kinetic energy, however their temperatures subside as well as their speeds slow down as well and become non-relativistic (hence the term "*cold*" dark matter in contrast to the other supposed kind, the relativistic *"hot"* dark matter, such as neutrinos and other particles) and are thus considered to be one of the most ideal candidates for being the constituent particles of the dark matter. Only a cold form of dark matter could have contributed to the process of structure formation leading to the cosmic structures we observe today.

However, in addition to that, axions seem to evolve in a non-linear manner following the onset of QCD transition and have additional effects from cosmological factors, such as the *Hubble* expansion. Hence, it is difficult to predict their behavior owing to a number of mathematical and cosmological considerations. Besides, the axion field produced as a result of symmetry breaking (PQ), undergoes time-dependent random fluctuations as the universe evolves, and thus, the field takes the form of an oscillating field around a mean value. An important stage in axion field evolution is the onset and duration of inflation, whether the PQ symmetry is broken before (or during) the onset of inflation or post-inflation, once a minimum has been reached in the inflation, i.e. the inflaton field has equilibrated to its minimum value in its potential valley, and the new episode of universe's evolution, the *Reheating*, begins.

Here, in this study, we contemplate the second scenario wherein the U(1) PQ symmetry breaks after the onset of inflation, while it has receded and the epoch of Reheating is commenced. As a result, the topological defects in the quantum vacuum produce a network of two-dimensional elementary objects, "Strings" and their corresponding "Domain Walls" [14], known as the "String-Wall" systems, which are formed at the onset of QCD transition and following the production of instantons. The formation of these systems and their subsequent collapse, later on through instantonic effects, in turn, produces several successive generations of cosmological axions, collectively forming a large number density of these particles, which later on become the relic cold dark matter in our universe [15]. This is the ad hoc hypothesis of axion production from instantonic effects that is so far the most plausible scenario of the origins of axions in our universe. However, an appropriate detection mechanism and the knowledge of the right mass range are important in order to detect these particles.

## 2. The Model:

The dynamics of a field evolving in space-time in an expanding universe are described by the Friedmann Equations, obtained from the Einstein Field Equations, using an appropriate metric. These equations laid down with the help of a spatially flat Friedmann-Lemaître–Robertson-Walker (FLRW) background, while incorporating the FLRW metric, can be expressed along with the metric as follows below [16]:

$$\frac{\ddot{a}_s(t)}{a_s(t)} = -\frac{4\pi G}{3}(\rho + 3p) + \frac{\Lambda}{3} \qquad 1.1a$$

$$ds^2 = -dt^2 + a_s(t)[dx^2 + dy^2 + dz^2] \qquad 1.1b$$

Where $a_s(t)$ is the scale factor, each dot depicts one-time differentiation with respect to the cosmic time, $G$ is the universal gravitational constant, $\rho$ and $p$ are the mass density and pressure of the fluid matter, respectively, and $\Lambda$ is the cosmological constant.

We consider bosonic fields arising out of the spontaneous broken PQ symmetry that acquire mass and become axion (or Axion Like Particles, ALP's) particles.

The interaction Lagrangian for these bosonic fields, often termed as the *"invisible axion"* particles, while ignoring the kinetic and other terms, is given by [3]:

$$\mathcal{L} = \frac{J^\mu \partial_\mu \varphi}{f_a} \qquad 1.2$$

Here $J^\mu$ is the associated Noether current of the broken PQ global symmetry, whereas $\varphi$ are the complex, pseudo-scalar bosonic fields (that become the axionic/ALP fields) and $f_a$ is the axion decay constant (a measure of axion-two-photon coupling, $g_{a\gamma\gamma}$), corresponding to the PQ symmetry breaking scale (also often written in the literature as $f_{PQ}$), and inversely proportional to the axion field's mass.

The axion field, $a(x)$, itself is the phase of the complex scalar field $\varphi$, described by:

$$\varphi(x) = \frac{1}{\sqrt{2}}(v + r(x))(e^{\frac{ia(x)}{v}}) \qquad 1.3$$

The radial mode of this field, shown here as the $r(x)$, is a massive field with mass $m_r$. At high temperatures, after the QCD transition takes place, the axions have mass acquired from instanton effects, albeit negligible. The QCD Phase transition (which sets the QCD temperature scale) involves the thermal transitory phase when the primordial quark and gluon condensate, created earlier in vacuum during the low-energy regime at zero temperature, is disappeared and the chiral symmetry is restored [6, 7].

The most significant element of the PQ theory and a decisive factor in the emergence of axions is the decay constant, $f_a$, which we shall limit our concentration on in this report.

The value of the axion decay constant is a free parameter in the theory under consideration here, with a broad range of mass values attributed to it. However, in view of physical and cosmological considerations, a (broad) realistic window of the coupling's possible range has been obtained as $3 \times 10^9 \geq f_a \leq 1 \times 10^{12}$ $(GeV)$ [4, 53], which approximately corresponds to an axion mass range of roughly ~$5.5 \mu eV \geq m_a \leq 3.0\ meV$ (with a somewhat lax lower bound, and conversely, a stringent bound on the upper limit). Hence, most of the axion searches are carried out within this parameter space.

The other main parameters in our model are the environmental factor, i.e. the misalignment angle, $\theta_i$, and the critical temperature ($T_c$) at which the transition takes place.

Following the PQ symmetry breaking, the axion is created with a spectrum of misalignment angles in various disconnected regions, however as the onset of inflation happens during the post–inflation era, the instanton effects cause the value of the misalignment angle to fall to a universal value oscillating in its minimum potential. The accepted value of this minimum is the root- mean square value of $\langle \vartheta_a \rangle \geq \frac{\pi}{\sqrt{3}} = 1.8137$, the lower bound around which the axion misalignment angle fluctuates (as existing in the current post-inflation epoch.) Once the involved calculations are carried out, this corresponds to the value of axion coupling around $f_a = 2.82 \times 10^{11}$ GeV, thus making an upper bound on the axion mass. On the other hand, the lower value of the coupling finds a lower limit of $f_a > 10^8$, following stringent astrophysical considerations.

Based upon earlier work, following a Dilute Gas Approximation (DGA) model conceived by Turner [44] (and further developed by Bae *et al.* [45]), assuming the axions to exist in the form of a dilute instanton gas following the onset of decoupling, the mass of axion/ALP's is described as follows for the dilute gas temperature Λ (in MeV):

$$m_a^2 = \frac{\alpha_a}{f_a^2 (T/\Lambda)^n} \Lambda^4 \qquad 1.4$$

The temperature at which the axions slow down and become the non-relativistic cold dark matter and undergo oscillations, as based upon the QCD equation of state, is found to be approximately equal to [46]:

$$T_c \cong 2.09 \left(\frac{f_a}{10^{10} GeV}\right)^{-0.1655} GeV \qquad 1.5$$

Corresponding to an upper limit of:

$$f_a \leq 3 \times 10^{17} GeV$$

However, in the post-inflation scenario, the coupling is endowed with string networks and domain walls production, and hence, it is restricted down to a lower value, as discussed earlier:

$$f_a \leq 10^{11}\ (GeV)$$

As per the misalignment mechanism [47], the value of the initial misalignment angle, $\theta_i$, spans the region:

$$\theta_i \in \left[\frac{1}{2}, \frac{\pi}{\sqrt{3}}\right] \qquad 1.6$$

Which produces the viable axion window in the post-inflation scenario of:

$$10^{-6} \leq m_a \leq 10^{-5}\ (eV) \qquad 1.7$$

In view of the measured non-absence of nEDM (Neutron Electric Dipole Moment) [52], an upper bound of $\theta_i < 10^{-10}$ is fixed for the value of the misalignment angle.

Two scenarios of PQ symmetry breaking and axion production are possible here and need to be considered. In the first scenario, the PQ symmetry breaks a priori to the epoch of inflation begins (called the *pre-inflation scenario*), while in the second one, the PQ symmetry breaks a posteriori when the epoch of inflation has already begun as the nascent universe evolves during the ensuing highly-accelerated inflation epoch, generally known as the *post-inflation schemes*.

The universal expansion (and consequently, the change in scale factor, which are generally gauged with the help of the Hubble parameter) has direct correlation with the temperature of the universe, which in turn affects the value of the axion potential [7].

As the universe temperature descends, the value of Hubble parameter also decreases, however the axion potential becomes deeper and its value increases. As per the equation of motion (obtained from the Dilute Instanton Gas model approximation [50]), the dynamics of the axion's field are governed by:

$$\ddot{a} + 3H\dot{a} + m_a^2(T)f_a \sin\left(\frac{a}{f_a}\right) = 0 \qquad 1.8$$

As argued in detail in [7], as the temperature cools down further the attractive axion potential reaches the expansion-driven Hubble friction. When the value of the temperature-dependent axion mass approaches nearly three times the value of the Hubble parameter, the axion field undergoes oscillation at the axion Compton frequency corresponding to the axion mass value. This is soon followed by axion number density becoming an adiabatic invariant and axion taking the form of cold dark matter. This happens assuming the scenario that the PQ symmetry breaks a priori to the epoch of inflation (the *pre-inflation scenario*), with the value of the misalignment angle $\vartheta$ settling down to a fixed place $\vartheta_0$ in the potential valley's bottom, as the nascent universe evolved during the highly-accelerated inflation epoch.

In the alternate scenario, the *post-inflation scheme*, PQ symmetry breaks following the onset of inflation, as the accelerated episode of inflation begins and the universe undergoes its massive accelerated expansion, and it is not possible to determine the value of $\theta$. The value, thus, has to be integrated numerically over all the possible values and an average value is chosen, which is typically $\theta_i \cong 2.1$. In this scenario, the free parameter of $\theta$ is eliminated, and one obtains instead of that the axion abundance $\Omega_a$ as a function of the decay constant $f_a$, as follows.

$$\Omega_a = 2.606 \left(\frac{n_a^*}{S^*}\right)\left(\frac{\Omega_\gamma}{T_\gamma}\right) m_a \qquad 1.9$$

Where the ratio $n_a^*/S^*$ is the ratio of post-axion oscillation values of the photon entropy density and the average axion density, respectively (and takes the value of a constant as soon as axion field undergoes oscillation) and the ratio $\Omega_\gamma/T_\gamma$ is the photon abundance's ratio to its temperature around the decoupling era (and as indirectly estimated today by the CMB).

The axion mass is bound within the μeV region owing to cosmological considerations (in order not to exceed the observed CDM density), which is known as the so-called "Anthropic window" [3]. This is, thus, the narrow window for ultra-light cold dark matter axion (and axion-like particle) searches.

The critical temperature ($T_c$) in the calculations is defined as the temperature when the axion field undergoes coherent oscillations around its minimum value (at the particular moment, which is known as the "critical time", or $t_c$):

$$m_a(T_c) \sim 3H(T_c) \qquad 1.10$$

Where, $m_a(T_c)$ is the mass of axionic field at the transition temperature and the Hubble parameter $H$[§1], as a function of the critical temperature, is given by a form of the Friedmann equation (with flat geometry and vanishing cosmological constant):

___________

§1 The value of the Hubble parameter and the Scaling factor are adopted in our model as below,

$$H_0 = 67.4 \pm 0.5 km.s^{-1}.MPc^{-1}, \quad h = \frac{H_0}{100} km.s^{-1}.MPc^{-1} \sim 0.7$$

$$H(T_c) = \sqrt{\frac{8\pi}{3M_p^2} \cdot \rho_c} \quad \text{1.11a}$$

$$= \sqrt{\frac{8\pi^3}{90} G g_c(T_c) \cdot T_c^2} \quad \text{1.11b}$$

$$= \sqrt{\frac{8\pi^3}{90} g_c(T_c) \cdot \frac{T_c^2}{M_p^2}} \quad \text{1.11c}$$

Here, $g_c$ is the relative degrees of freedom at the critical time t, G is the Newton's gravitational constant (which can be alternatively written in terms of the Planck mass, $M_p$), and $\rho_c$ is the critical energy density of the universe at the onset of the transition to the critical temperature.

We follow an approach, similar to Buschmann *et al.* [17] and Hiramatsu *et al.* [18, 19, 20], in constructing an axion model taking into account the three fundamental mechanisms for axion production, viz. misalignment [21, 22], global string decays [23] and domain wall decays [24], since all these three, especially the two principal mechanisms, the axion field misalignment mechanism and the decay of global axion strings and domain walls, equally contribute to post-inflation era axion production.

The energy density with the cosmological constant ($\Lambda$) and the Susceptibility ($\chi$) of topological charge at the temperature $T = 0$ are given by:

$$\rho_\Lambda = \frac{\Lambda M_p^2}{8\pi}, \quad \text{1.12a}$$

$$\chi(0) = m_a^2 f_a^2|_{T=0} = (75.6 MeV)^4 = (0.0245 fm)^{-4} \quad \text{1.12b}$$

which suppresses the temperature dependence and fixes the value of the zero temperature axion mass as a function of the coupling constant alone, and can be described by an expression [9]:

$$m_a(0) \cong \frac{5.691\,(2) \times 10^6\,GeV}{f_a} \,(eV) \quad \text{1.13}$$

## 3. Fitting of the Model to Cosmological Parameters

Similar to the earlier calculations by Wantz and Shellard [6], among others, we adopt a high-temperature cosmological model in an era soon after the decoupling, with a sufficiently high-temperature value of the decoupling temperature, around $T = 400$MeV (way above the QCD transition).

We carry out our calculations around the accepted value of axion density ($\Omega_a$) [25] with the express assumption that it is equal to the total cold dark matter density ($\Omega_{CDM}$):

$$\Omega_a h^2 = \Omega_{CDM} h^2 \quad \text{2.0}$$

Based upon the WMAP, BAO and Type 1a Supernova (SNe) data:

$$\Omega_{CDM} = 0.229 \pm 0.015 \qquad 2.1$$

On the other hand, the major component of the universe, i.e., the dark energy (or quintessence) density, $\Omega_\Lambda$, based upon a flat universe assumption, and using the Type 1 Supernovae data [26], is estimated at:

$$\Omega_\Lambda = 0.725 \pm 0.016 \qquad 2.2$$

In addition, it is also assumed in this model, similar to several earlier models, that the value of the radial mass ($m_r$) is on the order of the axion decay constant, $f_a$, well within the acceptance window of axion searches as per the cosmological observations and calculations.

We follow an axion string network based theoretical model similar to the approach by Kawasaki *et al.* [24], Hindmarsh *et al.* [27], Yamaguchi *et al.* [8], Vilenkin [14] and Shellard [30, 31], under the String theory framework, using results from their simulations, and work with the calculations and simulations similar in the form as set up by Gorghetto *et al.* [9]. However, no string network simulations are carried out, instead our calculations are setup around the results obtained from those simulations and accepted values of cosmological parameters as determined by the latest results of the Planck collaboration [25].

We attempt to fit our calculated results to a power law in order to estimate the value of $f_a$, as follows below. We consider three main contributions to the total cold dark matter density, i.e., the density as a result of axions radiated from the strings ($\Omega_a^{str}$), the ones radiated from the decay of string-wall systems ($\Omega_a^{dec}$), and the axions produced from the misalignment mechanism ($\Omega_a^{mis}$). The calculations for our model are carried out as follows, using a fitting routine working around a set of parameters, as given in the Table 1.

$$\Omega_a \sim \kappa f_A^\alpha \qquad 2.3a$$

$$\Omega_a h^2 = k \left(\frac{f_A}{GeV}\right)^\alpha \qquad 2.3b$$

$$\Omega_a^{tot} h^2 = \Omega_a^{str} h^2 + \Omega_a^{mis} h^2 + \Omega_a^{dec} h^2 \qquad 2.3c$$

$$\Omega_a^{str} h^2 = 3.63 \times 10^{-2} \times \left(\frac{N_{DW}^2}{1.0}\right) \times \left(\frac{\beta'}{51.04}\right) \times \left(\frac{g_{*,1}}{80}\right)^{-(n+2)/2(n+4)} \times \left(\frac{f_A}{10^{10} GeV}\right)^{(6+n)/(4+n)} \qquad 2.3d$$

$$\times \left(\frac{\Lambda_{QCD}}{400 MeV}\right)$$

$$\Omega_a^{mis} h^2 = 4.52 \times 10^{-4} \times \left(\frac{1.99}{c_{av}}\right) \times \left(\frac{g_{*,1}}{80}\right)^{-(n+2)/2(n+4)} \times \left(\frac{f_A}{10^{10} GeV}\right)^{(6+n)/(4+n)} \qquad 2.3e$$

$$\times \left(\frac{\Lambda_{QCD}}{400 MeV}\right)$$

$$\Omega_a^{dec} h^2 = 1.20 \times 10^{-3} \times \left(\frac{\beta_2}{62}\right)^{2/(4+n)} \times \left(\frac{g_{*,2}}{75}\right)^{-(n+2)/2(n+4)} \times \left(\frac{f_A}{10^{10} GeV}\right)^{(6+n)/(4+n)} \qquad 2.3f$$

$$\times \left(\frac{\Lambda_{QCD}}{400 MeV}\right)$$

Following the approach by several investigators before us, we consider it appropriate to adopt a value of the power n as the starting value before parametrizing it:

$$n = 3, \qquad 2.4$$

These parameters were fed into a fitting program based upon a multi-parameter space least square fitting scheme and a set of values was obtained for the required decay constant ($f_A$),

$n$ and alpha ($\alpha$), corresponding to the best known cosmological bounds as delineated in the beginning of this section.

As a result of our calculations and fitting the parameters to the current dark matter density, we obtain the values of α and κ as:

$$\alpha = 1.174, \kappa = 0.03799 \qquad 2.5$$

| Parameter | Value |
|---|---|
| $\Omega_a h^2$ | 0.120 |
| $\beta_1$ | 60 |
| $\beta'$ | 51.04 |
| $\beta_2$ | 62 |
| $g_{*,1}$ | 80 |
| $g_{*,2}$ | 75 |
| $\Lambda_{QCD}$ | 400MeV |
| $n$ | 3 |
| $\alpha$ | 1.16667-1.185 |
| $f_A$ | $4.0 \times 10^{10} - 8.0 \times 10^{11} GeV$ |

Table 1: The Parameter space for our model.

The theoretical value for α is usually obtained as a ratio $\frac{n+6}{n+4}$ =7/6 or 1.166667 in calculations carried out elsewhere, such as by Kawasaki and Hiramatsu *et al.* [18, 19], however, in our calculations here, the fitted value for alpha is obtained for the model, based upon our fitting routine, as 1.174. This is an important outcome of our calculations and a major differing factor from all the earlier calculations, albeit within the same window. We expect that it could possibly have significance in estimating the correct axion mass window, although these are not far from the values obtained by Kawasaki *et al.*, Buschmann et al., Hiramatsu *et al.* and others, they could be helpful in pinpointing a value for a cavity-based search where a fixed frequency search is sought in view of experimental limitations.

Thus, the final expression for the $\Omega_a$, using the above value of α, is obtained as:

$$\Omega_a h^2 = (3.799 \times 10^{-2}) \times \left(\frac{f_A}{10^{11} GeV}\right)^{1.174} \qquad 2.6$$

With the value of axion decay constant ($f_A$) obtained accordingly as:
$$f_A = 7.07 \, (\pm 0.47) \times 10^{11} GeV, \qquad 2.7$$

Which helps us obtain the corresponding axion mass value as:
$$m_a = 78.58 \, (\pm 5.0) \, \mu eV \qquad 2.8$$

The corresponding Compton frequency [32] for the suggested axion mass above is obtained appropriately, in a usual manner, as:
$$\nu_\alpha = 18.999 \, to \, 19.001 \, GHz \qquad 2.9$$

These are the results we obtained from our calculations and fitting routines, as based upon and modifying the pertinent calculations and simulations carried out elsewhere before us, making it feasible to devise a suitable experiment in a narrow-frequency range resonant cavity setting.

## 4. Experiment Design and Prospective Implementation

Based upon the short frequency range (falling within the Ku microwave band) obtained from our calculations, the next step is to implement this value and these ideas and probe this tangible range in an actual axion search experiment.

We adopt an approach similar to the one devised earlier by author [32], based upon the inverse of Primakoff effect [33], which was first discovered for pions, whereby an axion can interact with an ambient powerful magnetic field and convert into a pair of photons and the latter can be detected in an appropriate detection setting using any of various methods.

A review of various axion detection strategies based upon their electromagnetic detection can be found here [43].

The two-photon axionic decay width (under this model) is given by:

$$\Gamma_{A-\gamma\gamma} = \frac{g_{A\gamma\gamma}^2 m_A^3}{64\pi} = 1.1 \times 10^{-24} s^{-1} \left(\frac{m_A}{eV}\right)^2 \qquad 3.1$$

Figure 1.0 illustrates a magnetic field-induced resonant conversion of an axion into a microwave photon under the inverse Primakoff effect.

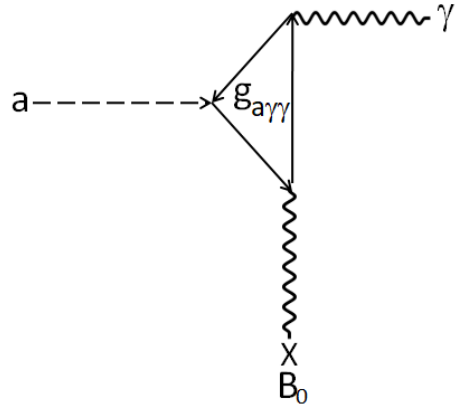

Figure 1.0: Magnetic field-mediated conversion of an axion into a photon under a process inverse of the Primakoff effect.

The interaction is supposedly carried out between an axion and a photon by means of the axion's two-photon vertex, as shown in the diagram. This vertex facilitates a possible interaction of axions with electromagnetic fields, especially in the presence of very strong magnetic fields, as described by the Lagrangian:

$$\mathcal{L}_{a\gamma\gamma} = \frac{1}{4} g_{a\gamma\gamma} F_{\mu\nu} \tilde{F}^{\mu\nu} \varphi_a \qquad 3.2$$

Here, $g_{a\gamma\gamma}$ is the dimensionless axion-two-photon coupling, $\varphi_a$ and $F_{\mu\nu}$ is the electromagnetic field tensor (with $\tilde{F}^{\mu\nu} = e^{\mu\nu\lambda\rho}F^{\lambda\rho}$ its dual, with $e^{\mu\nu\lambda\rho}$ being the Levi-Civita tensor with $e^{0123} = 1$).

In terms of the vector electric and magnetic fields, this becomes:
$$\mathcal{L}_{a\gamma\gamma} = g_{a\gamma\gamma}(\mathbf{E}.\mathbf{B})\varphi_a \qquad 3.3$$

The model-dependent value of the coupling $g_{a\gamma\gamma}$ is obtained as:
$$g_{a\gamma\gamma} = \frac{\alpha}{2\pi f_A}\left(\frac{E}{N} - 1.92(4)\right) = \left(0.203(3)\frac{E}{N} - 0.39(1)\right)\frac{m_A}{GeV^2} \qquad 3.4$$

The coupling depends upon the values of the electromagnetic and color anomalies, E and N, respectively, of the axion-associated axial current. We assume here $\frac{E}{N}=0$ under the Kim-Shifman-Vainshtein-Zakharov (KSVZ) model [34].

While following these lines, the probability of an axion produced in a strong magnetic field, as a result of inverse Primakoff effect, can be expressed as [32, 33],
$$P_{a\to\gamma} = \left(\frac{1}{2}g_{a\gamma}BL\right)^2\frac{\sin^2(\frac{qL}{2})}{(\frac{qL}{2})^2} \qquad 3.5$$

Where B is the intensity of the applied magnetic field, L is the cavity length and q is the momentum transfer [32].

The Signal-to-Noise-Ratio, an important parameter in the experimental detection of any signal, is given by the Dicke Radiometer equation, following the usual microwave signal measurement conventions, which is as given below [35].
$$S/N = \frac{P}{k_B T}\left(\frac{t}{\Delta f}\right)^{1/2}$$

Here, $k_B$ is the Boltzmann's constant and $t$ is the integration time.

Thus, in a nutshell, working under this theoretical framework, the possible detection of axion can be carried out, if these particles or such excitations exist in nature.

The central part of our detection scheme are a hi-finesse resonant RF cavity and an antenna weakly coupled to it, while this setup is mounted in the bore of a high-intensity magnetic field. The output of the antenna is coupled to an extremely sensitive Josephson Junction (JJ) based amplifier device that can amplify the extremely faint electromagnetic signal arising from any possible axion-photon conversion event. To be detected by the antenna and to be readout by a measurement system, a signal has to be of a sufficient level. Here, we introduce a new novel addition. A Tunneling Diode (Resonant Mode) just after the antenna and before the JPA stage further amplifies the detected signal, while in resonance.

| Parameter | Value |
|---|---|
| **Cavity Resonance Frequency (*f*)** | 17.00 to 18.99 GHz |
| **Pin-point Center Frequency (*f*$_c$)** | 18.999 GHz |
| **Cavity Length (*l*)** | Tunable, Variable |
| **Mean Operating Temperature (Cavity, T)** | 35.0mK |
| **Mean Magnetic Field (B)** | 9.0T (δB/B<1ppb) |
| **Integration Time** | 120-1200 s |
| **Transmon Central Frequency (ω/2π)** | 4.262 GHz |
| **Sensitivity (P)** | $10^{-16}$ W |
| **Mean Spectral Density (S)** | $10^{-18}$ W/Hz$^{1/2}$ |

Table 2: A summary of the cavity-based detection scheme parameters.

The particular JJ device reported here is a flux-driven Josephson junction Amplifier (JPA) design [36], implemented with the help of a commercial SQUID-coupled pumped Niobium JPA chip. In addition, as described in an earlier report [32], we employ a new Resonant Tunneling Diode [37, 38]-based amplification approach to further increase the sensitivity. We hope for a sensitivity up to around $10^{-18}$ W with a noise spectral density down to the order of $10^{-18}$ to $10^{-19}$ W/Hz$^{1/2}$. At a highly-uniform static magnetic field of approximately 8.0T < B < 10.0T with a high-resolution (around δT/T < 1ppb) and an operating temperature of ~35-55mK, and taking samples for sufficient longer times, we propose to achieve an operating window just above the Standard Quantum Limit and possibly detect any axion-induced signals within the cavity.

The pertinent parameters of the detection scheme are summarized in the Table 2. An image of a simulated response of cavity electromagnetic field obtained (from numerical methods) in a transverse slice of the cavity is shown in Figure 2.0, with a cartoon of the Primakoff conversion process superimposed onto that.

Figure 3.0 illustrates an overview of the central element of the detection scheme, i.e. a Josephson Junction, whereas the Figure 4.0 and Figure 5.0 illustrate block diagrams for the detection scheme and detection electronics for the aforementioned frequency range, with the respective cryogenic and room-temperature sections in an isolated environment individually marked, with especially the cryogenic section (the cavity and cryogenic amplification section) in a low-temperature mu-metal chamber isolated from the strong magnetic field. More details of the experiment, various components and relevant instrumentation may be found in our earlier report [32], generic details on microwave measurements can be found elsewhere, such as in [39].

Figure 6.0 illustrates a spectral plot of a test signal, a "false axion", injected into the cavity and detected with the amplification system, showing the measured Power Spectral Density ($W/\sqrt{Hz}$), as a function of frequency (measured in MHz.)

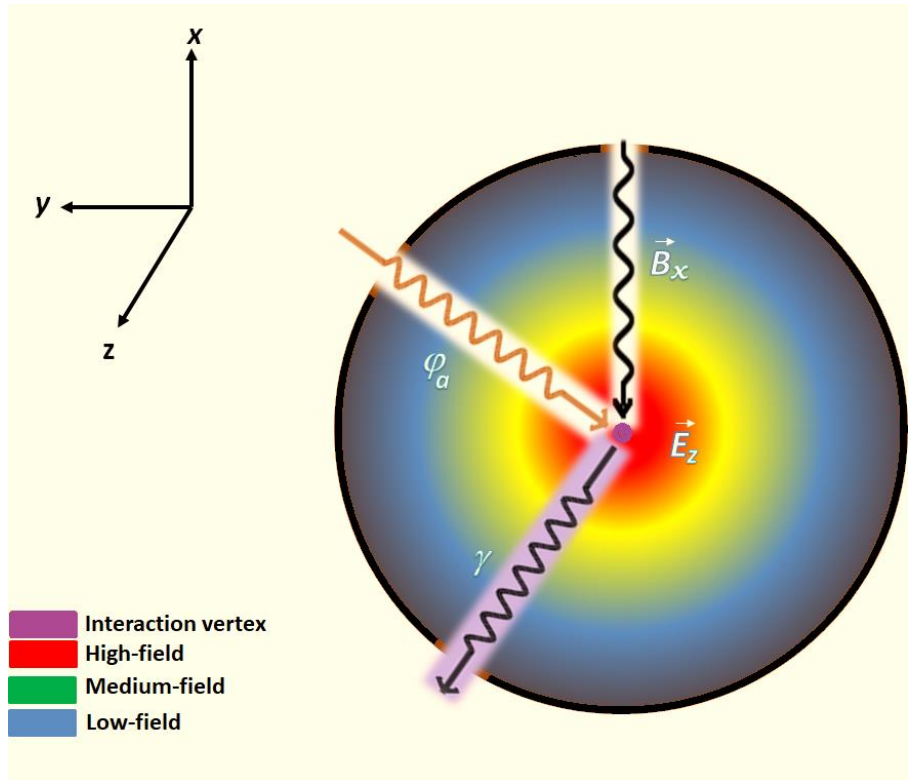

Figure 2.0: Cavity electromagnetic field distribution with a cartoon of an axion-photon conversion event facilitated by a strong magnetic field. Note that the magnetic field (radial) direction is perpendicular to the direction of the electric field (axial).

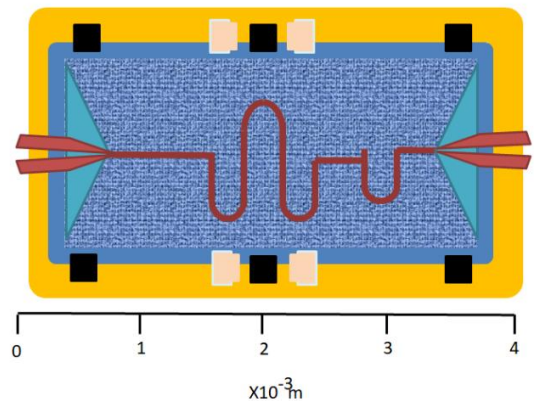

Figure 3.0: An overview of the superconducting Josephson Junction-based Parametric Amplifier (JPA), the central element of the detection scheme. The details are provided in the earlier report [32].

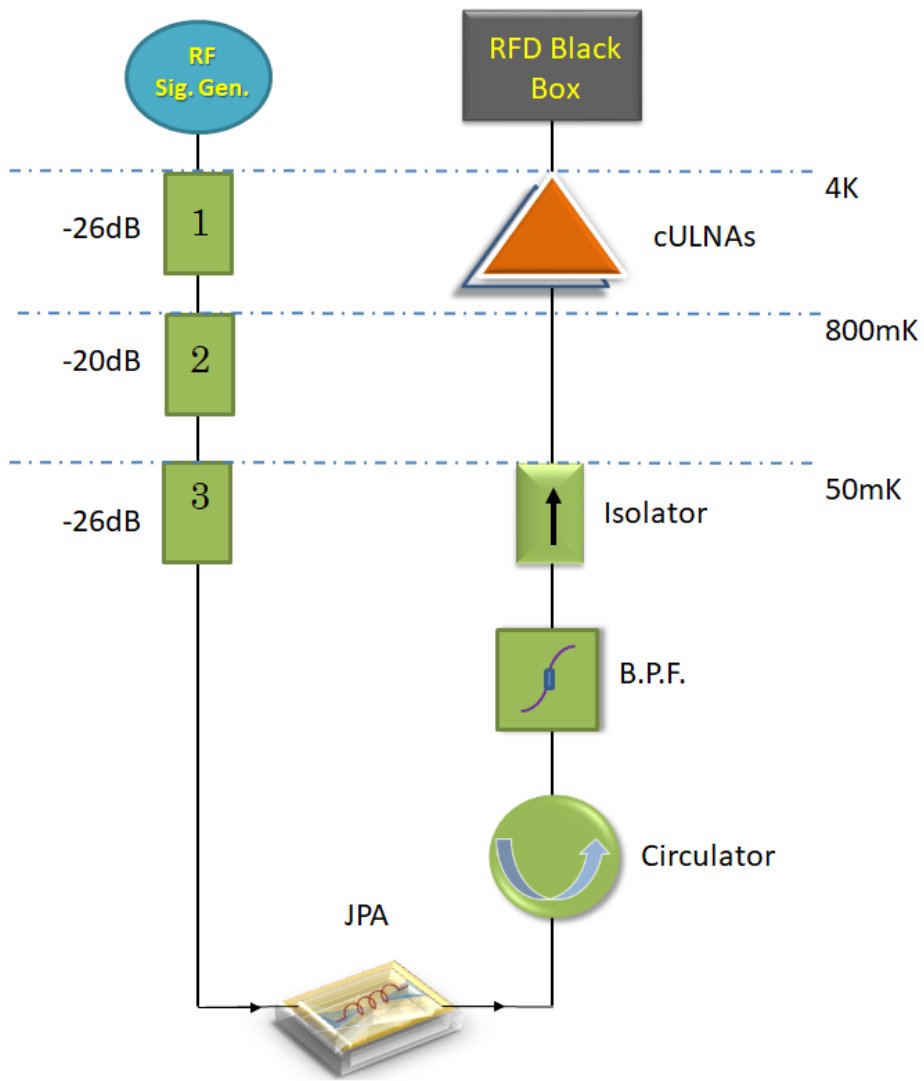

Figure 4.0: A block diagram of the detection components in the cryogenic measurement chain including the respective operating temperatures.

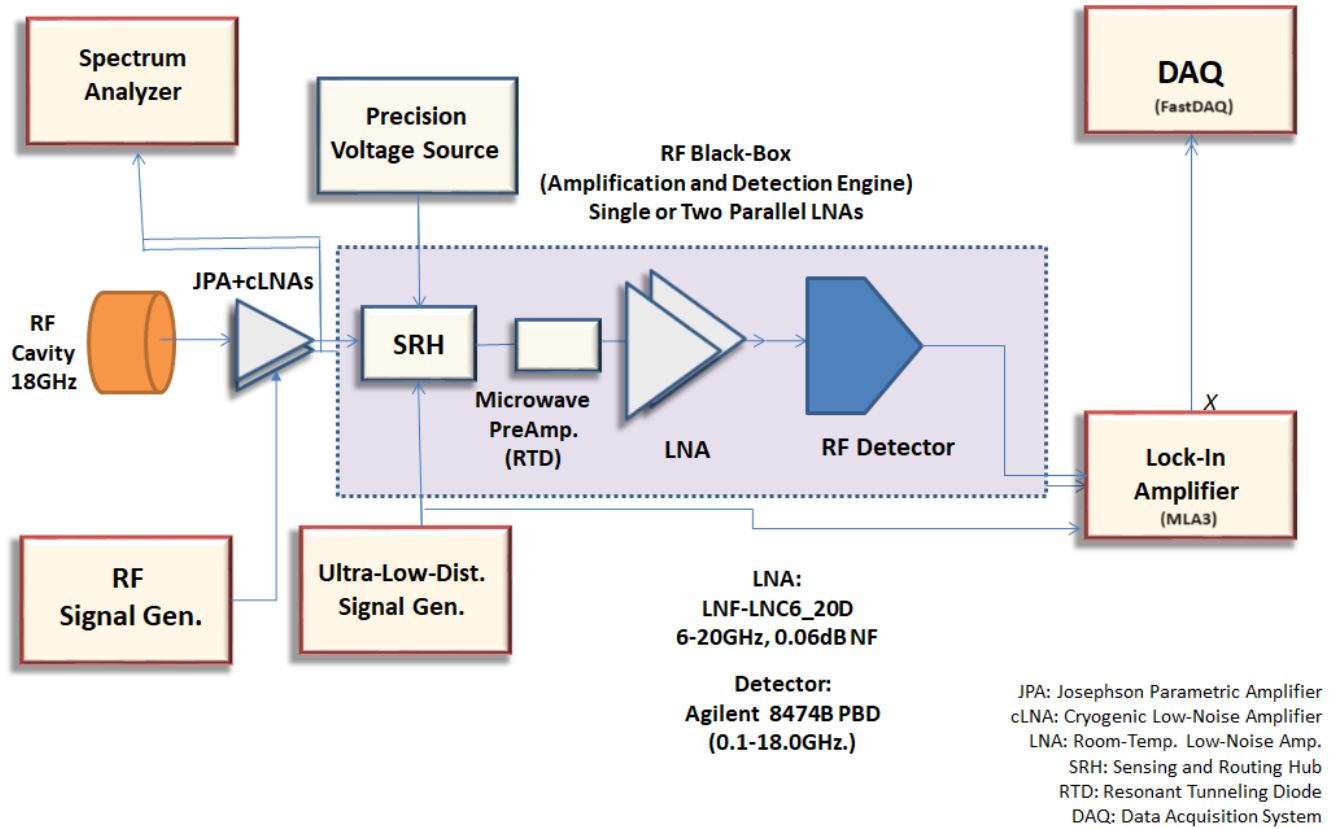

Figure 5.0: A block schematic of RF cavity, detection electronics, amplification and detection components and relevant instrumentation. As explained in detail in an earlier report [32], a weak electromagnetic signal in the cavity (on resonance) is weakly coupled to a fine gold-plated copper antenna and amplified by a combination of a JJ-based parametric amplifier and a set of cryogenic LNAs in parallel, all operating at cryogenic temperatures, till the signal is amplified sufficiently to be readout by the room-temperature electronics (operated at $10^{o}C$). A test signal for calibration is fed from an RF Signal Generator (1-10GHz). A home-made high-frequency Signal Sensing and Routing Hub (SRH), combines the detected signal with a carrier wave from an Ultra-Low-Distortion Signal Generator (1-100KHz), feeding it to a Resonant Tunneling Diode (RTD) stage. From RTD the signal is amplified by high-gain Low-Noise Amplifiers till detected by an RF detector.

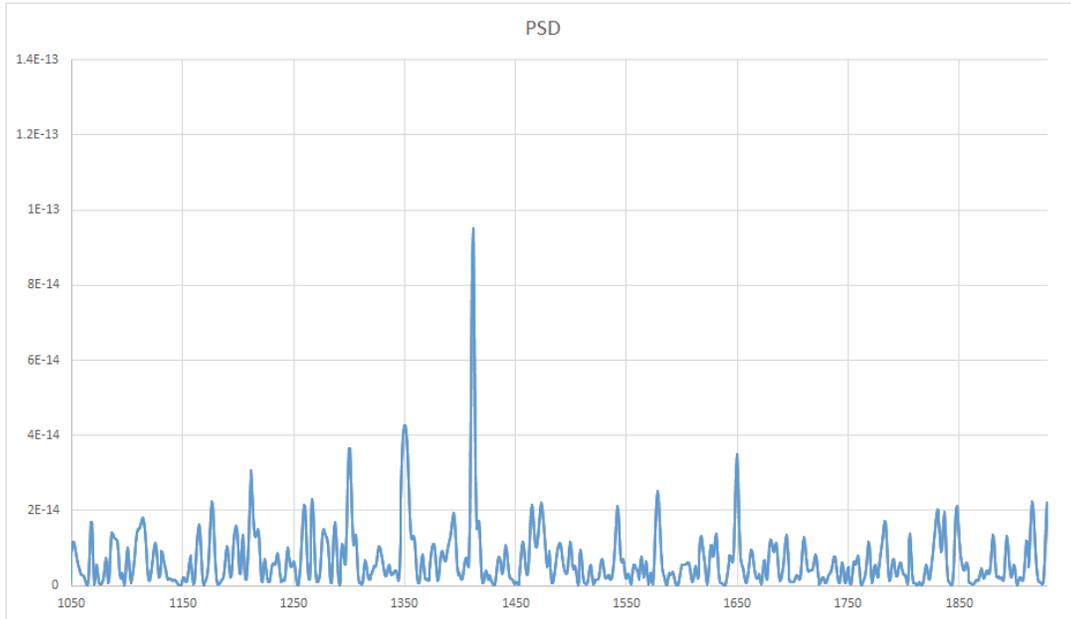

Figure 6.0: A plot of a test signal spectral response, as measured by the detection scheme, measuring an injected ultra-weak signal from an RF generator (Power Spectral Density in W/Hz$^{1/2}$ vs. Resonance Frequency MHz.)

## 5. Discussion and Conclusion

In this report we present some calculations of axion mass and coupling using standard methods and prescriptions. We assume that axions constitute the majority of cold dark matter content of the universe, a reasonable assumption in view of the current data, and consider three main contributions to axion production, which are the field misalignment mechanism, the decay of global axion strings and domain walls and the axions radiated from the strings. However, we consider the first two to be the major contributing factors to the total axion density in view of earlier reports. A viable axion mass window is obtained using our fitting routines and a method to probe this window is also suggested using an approach presented earlier. The value of the axion mass is quite plausible since it is very close and within the window of calculations and simulations carried out in recent times and also presents a specific center mass value for fixed resonance frequency value cavity-based searches.

In particular, the results of our calculations are on similar lines as the earlier recent simulations carried out by Borsanyi *et al.* [28], M. Gorghetto *et al.* [11] and Buschmann *et al.* [17, 51]. The axion mass range (and the corresponding frequency range) and the sensitivity of our experiment seem to be appropriate to detect an axion, subject to the existence of these particles. Recently, Kawasaki *et al.* [53] in their report has supported the value of $f_A \geq 10^{11}\ GeV$, coinciding with the decay constant $(f_A = 7.07\ (\pm 0.47) \times 10^{11} GeV)$, and, in turn, with the corresponding mass range, as proposed by us. On the other hand, Chang and Cui in their recent analysis of the dynamics of long-lived axion domain walls [54], have proposed a differing value of decay constant and mass window, one magnitude higher than those in our and other recent proposals. Kim et al. [29], while working on the similar lines of cosmic string networks formation in the post-inflation scenario, but with a new improvisation of the "Tetrahedralization of the Space", provide their own measures of axion abundance.

Working under a resonant cavity-based detection scheme, and making use of the Primakoff effect, as described in detail in one of our earlier reports [32], we have made some viable suggestions for a search for the axions/ALP's to be carried out at or around this frequency, which is a bit difficult due to practical considerations but not impossible to achieve in the contemporary era. It must be mentioned here that this is an extraordinary and remarkable frequency, just at the end of the microwave 'Ku' band.

An important concern is the limitations imposed upon the measurement of axion signal by quantum nature of the measurement. These include the Standard Quantum Limit, the Back-Action Noise by the detector and the quantum fluctuations inherent in the axion and electromagnetic quantum fields. A recent and quite pertinent discussion of these and the relevant parameters are available in the literature by Lasenby [40].

Axions have non-trivial significance in both physics and cosmology. They are predicted by the QCD, as discussed in the introduction of this report, as well as in the String theory [41]. Axions may also be radiated from Primordial Black Holes (PBH) as a result of their evaporation, as has been argued for some time, if PBH's can be found [48, 49].

Axion tomography can be carried out in association with gravitational waves detection [42] and their collective signatures could prove valuable in our understanding of the physical universe we inhabit.

**References**

1. Wilczek, F., Problem of Strong P and T Invariance in the Presence of Instantons, *Phys. Rev. Lett.* **40** (1978) 279.
2. Peccei, R. D. and Quinn, H. R., CP Conservation in the Presence of Instantons, *Phys. Rev. Lett.* **38** (1977) 1440.
3. Olive, K. A. *et al.,* Review of Particle Physics*, Chin. Phys.* **C38** (2014) 090001.
4. Dine, M., Fischler, W., The Not So Harmless Axion, *Phys. Lett.* **B120** (1983) 137; Preskill, J., Wise, M. B., Wilczek, F., Cosmology of the Invisible Axion, *Phys. Lett.* **120B** (1983) 127; Abbott, L. F., Sikivie, P., A Cosmological Bound on the Invisible Axion, Phys. Lett. **B120 (**1983) 133.
5. Seigar, M. S., The Dark Matter in the Universe, In Cold Dark Matter, Hot Dark Matter, and Their Alternatives, Morgan and Claypool: San Rafael, CA, USA, 2015.
6. Marsh, D. J. E., Axion cosmology, *Phys. Rept.*, **643** (2016), 1–79; Wantz, O. and Shellard, E., Axion Cosmology Revisited. *Phys. Rev. D* **82** (2010) 123508. [ArXiv:0910.1066].
7. di Cortona, G. Grilli, Hardy, E., Pardo Vega, J., Villadoro, G., The QCD axion, precisely. *JHEP* **01** (2016) 034.
8. Yamaguchi, M., Kawasaki, M., Yokoyama, J., Evolution of axionic strings and spectrum of axions radiated from them, *Phys. Rev. Lett.* **82** (1999) 4578 [ArXiv: hep-ph/9811311].
9. Gorghetto, M., Hardy, E. and Villadoro, G., Axions from strings: the Attractive Solution, *J. High Energ. Phys.* **07** (2018) 151 [ArXiv:1806.04677].

10. Gorghetto M. and Villadoro, G., Topological Susceptibility and QCD Axion Mass: QED and NNLO Corrections, *JHEP* **03** (2019) 33. [arXiv: 1812.01008].
11. Gorghetto, M., Hardy, E., Nicolaescu, H., Notari, A. and Reddi, M., Early vs. Late String Networks, *JHEP* **02** (2024) 223.
12. Gorghetto, M., Hardy, E. and Villadoro, G., More Axions from Strings, *SciPost Phys.* **10** (2021) 050 [ArXiv:2007.04990].
13. Venegas, M., Relic Density of Axion Dark Matter in Standard and Non-Standard Cosmological Scenarios, ArXiv (2021) [arXiv:2106.07796].
14. Vilenkin, A., Cosmic Strings and Domain Walls, *Phys. Rept.* **121** (1985) 263.
15. Davis, R. L., Cosmic Axions from Cosmic Strings, *Phys. Lett. B* **180** (1986) 225.
16. Garcia-Bellido, J., Cosmology and Astrophysics, *ArXiv* (2005) [arXiv:0502139].
17. Buschmann, M., Foster, J. W., Safdi, B. R., Early-Universe Simulations of the Cosmological Axion, *Phys. Rev. Lett.* **124** (2020) 161103.
18. Hiramatsu, T., Kawasaki, M., Sekiguchi, T., Yamaguchi, M., and Yokoyama, J., Improved estimation of radiated axions from cosmological axionic strings, *Phys. Rev.* **D83** (2011) 123531.
19. Hiramatsu, T., Kawasaki, M., Saikawa, K., Sekiguchi, T., Axion cosmology with long-lived domain walls, *JCAP* **1301** (2013) 001.
20. Hiramatsu, T., Kawasaki, Saikawa, K., and Sekiguchi, T., Production of dark matter axions from collapse of string-wall systems, *Phys. Rev.* **D85** (2012) 105020.
21. Co, R. T., Hall, L. J., and Harigaya, K., Kinetic Misalignment Mechanism, *Phys. Rev. Lett.* **124** (2020) 251802 [ArXiv: 1910.14152].
22. Chang, C.-F. and Cui, Y., New Perspectives on Axion Misalignment Mechanism, Phys. Rev. **D102** (2020) 015003 [ArXiv: 1911.11885].
23. Hagmann, C., *AIP Conference Proceedings* **1274** (2010) 103.
24. Kawasaki, M., Saikawa K. and Sekiguchi, T., Axion Dark Matter from Topological Defects, *Phys. Rev.* **D91** (2015) 065014. [arXiv:1412.0789].
25. Ade, P. A., Aghanim, N., Arnaud, M., Ashdown, M., Aumont, J., Baccigalupi, C., Banday, A. J., Barreiro, R. B., Bartlett, J. G., Bartolo, N. *et al.*, [Planck Collaboration] Planck 2015 results-xiii. Cosmological parameters. *Astron. Astrophys.* **594** (2016) A13.
26. Read, J. I., The local dark matter density, *J. Phys. G: Nucl. Part. Phys.* **41** (2014) 063101.
27. Hindmarsh, M., Lizarraga, J., Lopez-Eiguren, A., Urrestilla, J., Approach to Scaling in Axion String Networks, *Phys. Rev*. **D103** (2021) 103534.
28. Borsanyi S. *et al.*, Calculation of the axion mass based on high-temperature lattice quantum chromodynamics, *Nature* **69** (2016) 539.
29. Kim H., Park, J., Son M., Axion Dark Matter from Cosmic String Network, *ArXiv* (2024). arXiv:2402.00741.
30. Shellard, E. P. S., Strong Network Evolution. In: Davis A.C., Brandenberger R. (Eds.). Formation and Interaction of Topological Defects. 1995, NATO ASI Series, Vol. 349, Springer, Boston, MA.
31. Wantz, O., Shellard, E. P. S., Axion Cosmology Revisited, *Phys. Rev.* **D82** (2010) 123508.
32. Bukhari, M. H. S., A Table-Top Pilot Experiment for Narrow Mass Range Light Cold Dark Matter Particle Searches, *Universe* **6** (2020) 28.
33. Kahn, Y., Safdi, B. R., Thaler, J., Broadband and resonant approaches to axion dark matter detection, *Phys. Rev. Lett*. **117** (2016) 141801.

34. Tanabashi, M., Hagiwara, K., Hikasa, K., Nakamura, K., Sumino, Y., Takahashi, F., Tanaka, J., Agashe, K., Aielli, G., Amsler, C. *et al.*, Review of Particle Physics, *Phys. Rev.* **D98** (2018) 030001.
35. Dicke, R. H., The measurement of thermal radiation at microwave frequencies, In Classics in Radio Astronomy, Springer: Dordrecht, The Netherlands, 1946, 106.
36. Roy, A., Devoret, M., Introduction to parametric amplification of quantum signals with Josephson circuits, *C. R. Phys.* **17** (2016) 740.
37. Doychinov, V., Steenson, D. P., Patel, H., Resonant-Tunneling Diode Based Reflection Amplifier, In Proceedings of the 22nd European Workshop on Heterostructure Technology (HETECH), Glasgow, UK, 9–11 September (2013).
38. Sollner, T. C. L. G., Le, H. Q., Brown, E. L., Microwave and Millimeter-Wave Resonant Tunneling Devices, Technical Report for Electronics and Electrical Engineering, NASA: Lexington, MA, USA, January 1988.
39. Feng X., *et al.*, 6 to 26 GHz Detectors for High Data Rate ASK Signal Demodulation, Microwave Journal, September 15th (2014) pp. 1-9.
40. Lasenby, R., Parametrics of electromagnetic searches for axion dark matter, *Phys. Rev. D* **103** (2021) 075007.
41. Svrcek, P., Witten, E., Axions in string theory, *J. High Energ. Phys.* **2016** (2016) 051.
42. Gouttenoire, Y., Servant, G., Simakachorn P., Kination cosmology from scalar fields and gravitational-wave signatures, arXiv (2021) [arXiv:2111.01150].
43. Kim, J. E., Carosi, G., Axions and the strong CP problem, *Rev. Mod. Phys.* **82** (2010) 557. [arXiv:0807.3125]
44. Turner, M., Cosmic and local mass density of invisible axions, *Phys. Rev.* **D33** (1986) 889.
45. Bae, K., Huh, J., Kim, J., Update of axion CDM energy, *JCAP* **0809** (2008) 005.
46. Ballesteros, G., Redondo, J., Ringwald, A., Tamarit, C., Standard model-Axion-Seesaw-Higgs Portal Inflation, Five problems of particle physics and cosmology solved in one stroke, arXiv (2016) [arXiv:1610.01639].
47. Chang, C-F. and Cui, Y., New perspectives on axion misalignment mechanism, arXiv (2019) [arXiv:1911.11885].
48. Khlopov, M. Y., Primordial Black Holes, *Res. Astron. Astrophys.* **10** (2010) 495.
49. Bernal N. *et al.*, Axion dark matter in the time of primordial black holes, *Phys. Rev.* **D104** (2021) 075007.
50. Borsanyi S. *et al.*, Axion cosmology, lattice QCD and the dilute instanton gas, arXiv (2015) [arXiv:1508.06917].
51. Buschmann, M., Foster, J. W., Hook, A. *et al.*, Dark matter from axion strings with adaptive mesh refinement, *Nat. Commun.* **13** (2022) 1049.
52. Abel C. *et al.*, Measurement of the permanent electric dipole moment of the neutron, *Phys. Rev. Lett.* **124** (2020) 081803.
53. Kawasaki, M., Sonomoto, E. and Yanagida, T. T., Cosmologically allowed regions for the axion decay constant Fa, *Phys. Lett. B.* **782** (2018) 181 [arXiv: 1801.07409].
54. Chang, C-F. and Cui, Y., Dynamics of Long-lived Axion Domain Walls and Its Cosmological Implications, arXiv (2023) [arXiv: 2309.15920].